# Adversarial-Aware Deep Learning System based on a Secondary Classical Machine Learning Verification Approach

Mohammed Alkhowaiter [1,2], Hisham Kholidy [3], Mnassar Alyami [1], Abdulmajeed Alghamdi [1], and Cliff Zou [1,*]

1. College of Engineering and Computer Science, University of Central Florida, USA
2. College of Computer Engineering and Science, Prince Sattam bin Abdulaziz University, Saudi Arabia
3. College of Engineering, SUNY Polytechnic Institute, Utica, USA
* Correspondence: changchun.zou@ucf.edu; Tel.: +1-407-823-5015

**Abstract:** Deep learning models have been used in creating various effective image classification applications. However, they are vulnerable to adversarial attacks that seek to misguide the models into predicting incorrect classes. Our study of major adversarial attack models shows that they all specifically target and exploit the neural networking structures in their designs. This understanding makes us develop a hypothesis that most classical machine learning models, such as Random Forest (RF), are immune to adversarial attack models because they do not rely on neural network design at all. Our experimental study of classical machine learning models against popular adversarial attacks supports this hypothesis. Based on this hypothesis, we propose a new adversarial-aware deep learning system by using a classical machine learning model as the secondary verification system to complement the primary deep learning model in image classification. Although the secondary classical machine learning model has less accurate output, it is only used for verification purposes, which does not impact the output accuracy of the primary deep learning model, and at the same time, can effectively detect an adversarial attack when a clear mismatch occurs. Our experiments based on CIFAR-100 dataset show that our proposed approach outperforms current state-of-the-art adversarial defense systems.

**Keywords:** Computer security; Deep neural networks; Image forensics; Adversarial machine learning; Image manipulation detection;

## 1. Introduction

As machine learning (ML) technology, especially deep learning technique, in computer vision continues to advance, the challenges of adversarial attacks are becoming increasingly prevalent. Adversarial attacks refer to image manipulation to deceive computer vision tasks where the image seems correct at human perception [1]. Some of these attacks can lead to harmful failures in sensitive computer vision-based applications, such as targeting autonomous vehicles to mislead the AI system in those vehicles to recognize the road STOP sign as SPEED LIMIT 65. The increased demand for AI applications may increase the risks of this technology if it is not secured well before it is put on the market. Therefore, in recent years researchers have continued to develop algorithms and systems to prevent adversarial attacks. In this paper, we develop a novel adversarial-aware deep learning system by employing a classical ML algorithm as the auxiliary verification approach.

Deep Neural Network (DNN) theory, also called deep learning, accelerates the development of computer vision applications to advance [2–5]. Unlike other AI approaches, it can quickly learn complex patterns and representations from large and high-dimensional datasets. Therefore, according to Stone [6] study, DNN technology will be used in an expanding range of real-world applications within the next decade. Examples of these applications include autonomous vehicles, security surveillance cameras, and healthcare. However, this technology faces serious security challenges because of two factors. One is the high dimension and complexity of the input data to DNN models, which means it is hard to catch all potential attacks as adversarial attackers can insert small but enough perturbations to mislead the system. Second is the non-linearity in the decision boundaries of DNNs, resulting in unexpected and complex behaviors that are hard to predict.

*1.1. Inspiration*

In this paper, our proposed new idea in defending against adversarial attacks is inspired by analyzing communication war in the real world, as described below. Suppose in a war scenario or simulation; the Blue team uses satellite communication to operate its military. If the other side, the Red team, somehow is capable of modifying the Blue team's satellite communication without being detected, then the Blue team will be misled and could lose the war eventually. In defending against such an attack in disruption of its communication, the Blue team could add a radiotelegraphy secondary system to complement its main satellite communication because the radiotelegraphy, relying on a completely different mechanism, cannot be disrupted by the Red team satellite attack methodology. Although radiotelegraphy using Morse code has very limited bandwidth, it can transmit summary data that matches the complete data transmitted via satellite communication. In this system, if the receiver finds out that the information between the radiotelegraphy and the satellite communication do not match, it can tell that a Red team satellite-based attack is ongoing and thus will not be fooled by the misinformation.

Our proposed defense system against adversarial attacks uses the same philosophy as the war scenario described above. The deep learning image classification system is an analogy to satellite communication, which could be compromised by various adversarial attacks. However, we propose to use a traditional ML algorithm, such as RF, in analogy to radiotelegraphy, as the secondary verification system. Although it is less accurate than a normal deep learning image classification system, it is immune to most known adversarial attacks because it does not rely on neural network structure. In this way, we can detect adversarial attacks easily when there is a mismatch between the outputs of the primary deep learning module and the secondary RF module.

*1.2. Research Contributions and Paper Outline*

Based on a large-scale experiment and investigation, we find a ground similarity between various adversarial attacks on different deep learning models, which motivated us to develop this research work, as illustrated in Section 2.1. Our main contribution to this paper is integrating the primary deep learning model with an auxiliary traditional ML model that is not based on neural network architecture (presented in Section 2). Additionally, a new defense metric for selecting the highest Top_k predicted class probabilities of an input sample is introduced in Section 2.3. The misclassification issue of DNN models is also addressed in the same section, and an overall DNN model with improved accuracy is discussed.

Next, our method surpasses all other state-of-art defense methods in detecting multiple adversarial attacks using the CIFAR-100 dataset [7], which is shown in Section 3. A thorough discussion of our research is presented in Section 4, which covers both the solutions, the challenges, and the limitations encountered. Finally, a comprehensive conclusion is reached in Section 5, and potential future research avenues are identified to improve the reliability of adversarial detection models further.

*1.3. Related Work*

In this section, we briefly review state-of-the-art existing works on adversarial attacks and defenses. Also, we study the competitive detector methods that we compare our work with. These models are DkNN [8], LID [9], Mahalanibis [10], and NNIF [11].

1.3.1. Adversarial Attacks

In the past few years, many adversarial attacks have been proposed, and one and most common attack proposed by [12] is called The Fast Gradient Sign Method (FGSM). This attack adds a small perturbation to the target image in the direction of the gradient of the loss function with respect to the human-perception content in order to misclassify the trained targeted model. It is a white-box attack where the attacker fully knows the deep learning model, including its architecture, parameters, and training data. Later, a more efficient attack known as Deepfool [13] finds the smallest perturbation necessary to cause a DNN to misclassify an input image, which increases the attack success rate compared to FGSM.

The potential of deceiving the DNN models increased significantly with the past few years of adversarial attack development. Today, imperceptible perturbations can be added to input images with the flexibility of adjusting the attack goal to either a white-box or a black-box such as the one proposed in [14] and known after the founders' names Carlini & Wagner (CW) attack. This attack uses an optimization algorithm in order to find the smallest perturbation that minimizes a loss function that balances the size of the perturbation with the misclassification success rate. Moreover, the attack has the ability to incorporate constraints on the perturbation, like limiting the magnitude of the perturbation or restricting pixel values of the perturbed image. The power of this attack raises the challenge of defense solutions against multiple attacks at once.

The white-box approach becomes more desirable for adversaries as it was introduced by [15] and is known as the complete white-box adversary. The researchers found that the projected gradient descent (PGD) can lift any constraints on the amount of time and effort the attacker can put into finding the best attack. The iterative feature of the PGD attack makes it more effective than other attacks, such as FGSM, in finding imperceptible adversarial examples. The variety and effectiveness of adversarial attacks open a wide range of areas for researchers to develop different attacks, such as in [16,17], and finding defense mechanisms on the other side.

1.3.2. Adversarial Defenses

Authors of [18] categorize the adversarial defense mechanisms in computer vision into three approaches. The first approach targets the deep learning model by making modifications to the model itself in order to make it more resistant to adversarial attacks. The approach was initially employed by researchers Szegedy and Goodfellow [1,19] in 2013 and 2014, respectively. Years later, Madry [15] delved deeper into this approach by studying the robustness of neural networks against adversarial attacks from a theoretical standpoint, using robust optimization techniques. Despite its limitations, as discussed in the article by in [15], adversarial training has garnered considerable attention from the research community. In the paper [20], a new defense algorithm called Misclassification Aware adveRsarial Training (MART) is proposed. It distinguishes between misclassified and correctly classified examples during the training process. In another study [21], researchers suggest using dropout scheduling to enhance the efficiency of adversarial training when employing single-step methods. Researchers in [22] proposed a self-supervised adversarial training method, while [23] analyzed adversarial training for self-supervision by incorporating it into pretraining.

The second approach is a defense that targets the inputs to the model by cleaning inputs to make them benign for the target model. [24] proposed ComDefend that consists of a compression convolutional neural network (ComCNN) and a reconstruction convolutional neural network (RecCNN). The ComCNN model compresses the input image to maintain the original image structure information and purify any added perturbation. The RecCNN model, on the other hand, reconstructs the output of ComCNN to a high quality. This approach achieved high accuracy in defending multiple adversarial attacks. GANs architecture is another technique of input transformation introduced by [25]. Their method, Defense-GAN, learns the distribution of clean images. In other words, it generates an output image close to the input image without containing the potential adversarial perturbation.

The third approach is a defense of adding external modules (mainly detectors) to the target model. Among adversarial defense/detection techniques, [8] inserted a K Nearest

Neighbours model (*k*-NN) at every layer of the pre-trained DNN model to estimate better prediction, confidence, and credibility for a given test sample. Afterward, a calibration dataset was used to compute the nonconformity of every test sample for a specific label j. This involved counting the number of nearest neighbors along the DNN layer that differed from the chosen label j. The researchers discovered that in cases where an adversarial attack was launched on a test sample, the true label exhibited less similarity with the *k*-NN labels derived from the DNN activations across the layers.

The research in [9] characterized the properties of regions named adversarial subspaces by focusing on the dimensional properties vie using the Local Intrinsic Dimensionality (LID). The LID method evaluates the space-filling capability of the area around a reference by analyzing the distance between the sample and its neighboring points. A classifier trained using a dataset comprising three types of examples: adversarial as a positive class, normal and noisy (non-adversarial) as a negative class. The features of each sample associated with each category were then constructed using the LID score calculated at every DNN layer. Finally, a Logistic Regression (LR) model was fitted on the LID features for the adversarial detection task.

Researchers in [10] developed generative classifiers that could detect adversarial examples by utilizing DNN activations from every layer of the training set. They used a confidence score that relied on Mahalanobis distance. First, they found the mean and covariance of activations for each class and layer. Then, they measured the Mahalanobis distance between a test sample and its nearest class-conditional Gaussian using Gaussian distributions. These distances served as features to train a logistic regression classifier. The authors found that compared to using the Euclidean distance employed in [9], the Mahalanobis distance was significantly more effective in detecting adversarial examples and resulted in improved detection results.

In a study by Cohen et al. [11], they utilized an influence function to create an external adversarial detector. This function calculates how much each training sample affects the validation data, resulting in sample influence scores. Using these scores, they identified the most supportive training instances for the validation samples. To compute a ranking of the supportive training samples, a *k*-NN model is also fitted on the model activations. According to their claims, supportive samples are highly correlated with the nearest neighbors of clean test samples, whereas weak correlations were found for adversarial inputs.

## 2. Materials and Methods

This section introduces our proposed detection method in depth. It starts with the motivation, which lights up our research ideas. Then we introduce our model in detail. After that, we present the adaptive design for our defense method based on an application specific security goals.

### 2.1. Motivation and Threat Model

**Motivation:** After multiple assessments of the different adversarial attacks on different DNN models, we notice that once the attack succeeds on one deep learning model, it succeeds on other models as well, as shown in Table 1 that is obtained by running multiple adversarial attacks FGSM, Deepfool, CW, and PGD on ResNet-34 [26] as a target model using CIFAR-100 dataset. The generated adversarial samples are then tested on VGG16 [27] and DenseNet [28] DNN models classifications. We find out that the accuracy of the targeted model is almost similar to un-targeted DNN models. Researchers in [29] addressed the same issue, naming it "transferability" of adversarial examples, meaning that the generated samples from adversarial attacks on one targeted DNN model may work on different un-targeted DNN models. Therefore, a model from a different approach is interesting to be studied, and we selected Random Forest (RF) [30] decision tree-based classifier model for our study, considering all the challenges of using this limited model for image classification.

**Table 1.** Accuracy comparison of different DNN models before and after adversarial attacks on CIFAR-100 dataset.

|  | Attacks | Targeted model | Un-targeted models | |
|---|---|---|---|---|
|  |  | ResNet-34 | VGG16 | DenseNet |
| Accuracies (%) | without attack | 77.47 | 72.25 | 78.69 |
|  | FGSM | 34.25 | 35.09 | 36.19 |
|  | Deepfool | 25.78 | 24.79 | 24.84 |
|  | CW | 25.77 | 24.49 | 25.0 |
|  | PGD | 22.58 | 22.87 | 22.7 |

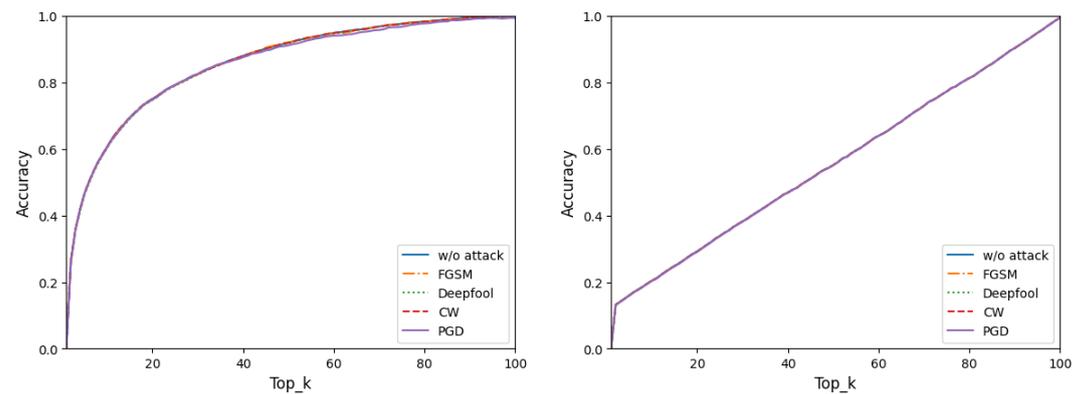

**Figure 1.** Classification accuracy over Top_k before and after different adversarial attacks using CIFAR-100 dataset, by two classical ML models: (a) Random Forest model, and (b) $k$-NN model. The accuracy under different adversarial attacks are almost identical, thus those resulting curves override each other and make a single purple color curve.

**Threat model:** Our threat model assumes that the attacker knows there is a detection method employed but does not know what is the detection method. In this setting, only the DNN model and its parameters are known to the adversary.

### 2.2. Proposed Methodology

We introduce our proposed adversarial attack detection method in this section. Our primary image classification system, shown in Figure 2, is based on the DNN approach, and we choose ResNet with 34 layers here for our investigation. The primary model could be any other DNN model that uses backpropagation because adversarial attacks exploit backpropagation to optimize the perturbations introduced to the input data on DNN models. The input is an image that could be a real image with no alteration or an adversarial generated sample from one of four attacks: FGSM, Deepfool, CW, or PGD. Our output of ResNet-34 is the highest probability index that indicates the class the image belongs to, which is referred to as Top_1 classification.

Unlike the primary approach, we use the classical ML model, Random Forest (RF) model, as a secondary model for adversarial attack detection. The RF model is a decision tree module based used in regression and multi-classification problems [31–35]. It is an extension of the bagging method as it utilizes both bagging and random feature selection to create an uncorrelated forest of decision trees. Also, it reduces overfitting and increases the diversity of the trees in the forest. The randomness in selecting the features for each tree determines and eliminates the inserted perturbations information on the adversarial samples as illustrated in Figure 1 where the accuracies of RF model before and after different attacks are almost identical. In the same figure, $k$-NN model is demonstrated as a classical ML model that is not affected by the added perturbations as well.

Our outputs of RF are the top k indices (Top_k) of the predicted class probabilities for the inputs. We selected Top_k and relied on it for our study to match the accuracy of the RF with the selected DNN model on the CIFAR-100 dataset that has 100 classes. Top_1 in the RF represents the worst accuracy as illustrated in Figure 1 whereas Top_100 represents 100 percent accuracy because its decision is always correct where the decision group includes all possible classes. When the parameter *k* in Top_k equals 22, the accuracy reaches around 77 percent, the same percentage as the primary DNN method prediction accuracy at the top_1 classification. Moreover, by adjusting the value of *k* on the Top_k classification, our methodology provides more control to its users and choices of selecting optimal security versus classification accuracy based on the AI application design, as it will be described in depth in section 2.3.

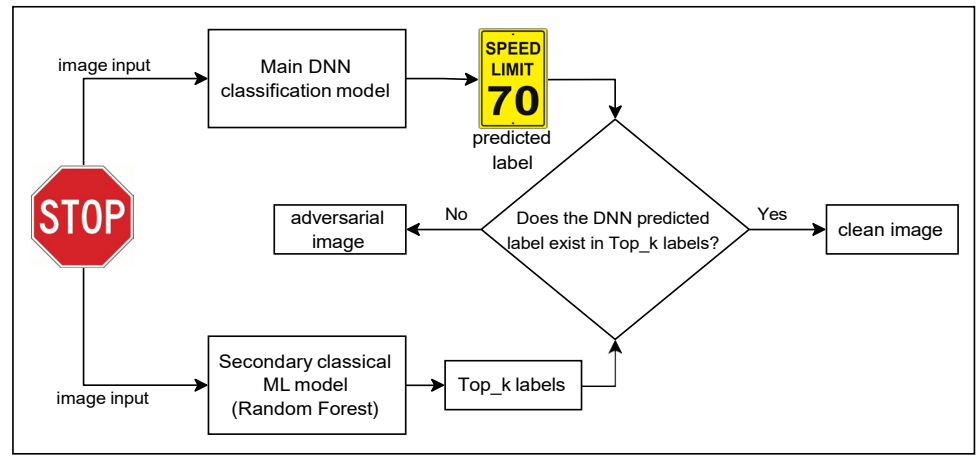

**Figure 2.** Proposed adversarial detection system design, which is composed of a primary DNN classification decision model and a secondary classical ML model for adversarial attack detection and verification.

2.2.1. Category of Image Dataset

Under Adversarial Machine Learning (AML), we run each adversarial attack individually on the DNN model, ResNet-34, using the test set in CIFAR-100 dataset, which contains 10K images. The attack success ratio of each one of the adversarial attacks varies, as illustrated in Table 1. During the categorization process as it is represented in Algorithm 1, each image x is first checked by DNN model for the correct label. The mispredicted result from DNN(x) adds x to SETmis set directly. In contrast, the correct predicted of x passes to AML(x) algorithm for a trial (e.g., FGSM), and the successfully applied attack output is added to SETadv set. The unsuccessful attack moved x to SETcrc set. In summary, we categorize the outputs into three sets as follows:

- SETcrc: The set of images that DNN can correctly identify by DNN.
- SETmis: The set of images that DNN misidentifies (misclassification).
- SETadv: The set of images produced by AML can successfully and deliberately make DNN misidentify as another object the attacker wants.

The percentage of misclassified images (SETmis) is maintained at 22.54 across various attacks. However, the percentages for the other categories vary depending on the strength of each attack and its parameters. Generally, the adversarial generated samples SETadv or the attack success ratio received the highest percentage among other sets in all four adversarial attacks.

2.2.2. Detection Algorithm

The adversarial image detection model, denoted as $Adv - aware(x)$, is addressed in Algorithm 2. We first pass a test image *x* to the primary DNN model, which is DNN(x) with *Top*_1, and to the secondary model, which is RF model with *Top_k* donated by RF(x, k). Then, we will have two outputs: a single-class prediction from the primary DNN model,

---

**Algorithm 1** Categorize Image Dataset.

---

**[SETcrc, SETmis, SETadv] = Category (Image dataset, DNN classification results)**
**Input:** {*x*, *right label*} ∈ CIFAR-100(*test set*), *DNN model DNN*(*x*), *adv_attack AML*(*x*)
**Output:** The three categories of image dataset according to DNN model classification and AML results.

Initialize SETcrc, SETmis, SETadv to be all empty

**for** image $x \in$ CIFAR-100 **do**
  **if** $DNN(x)$ is mispredict **then**
    SETmis ← *x*
  **else**
    **if** $DNN(x)$ is correct and AML(*x*) fail **then**
      SETcrc ← *x*
    **else**
      SETadv ← *x*
    **end if**
  **end if**
**end for**
**return** [SETcrc, SETmis, SETadv]

---

*y*, and *k* classes prediction from the secondary model *Top_k* as a list of *k* classes. We check whether *y* predicted class exists in the Top_k prediction list. If *y* exists in Top_k, then it returns a boolean "false" value for forged status with the DNN(x) label, *y*. Otherwise, it returns "true" without a label or None, which detects a possible adversarial sample.

For instance, from Figure 2, we use a *STOP* road sign as an input sample to our model. It passes to the primary model and the secondary model concurrently. In an adversarial attack scenario where the *STOP* sign image is a manipulated image, the predicted class from the primary model is *SPEED LIMIT 70*, whereas the second model provides Top_3 list of predictions, for example, *STOP*, *Roundabout*, and *No entry*. Our model will detect the input image as a forged "true" since the predicted class from the primary does not exist in the list of the secondary model. In the case of clean detection, the predictions have to be found in both model predictions. During our evaluation, we excluded misclassification samples in this section, and will be tackled in section 2.3.

*2.3. Defense System Adaptive Design*

This section discusses a new technique for selecting the best value of *k* in the Top_k used in the secondary model based on the underlying applications' specific requirements in terms of accuracy and security. Some applications require zero tolerance for attack success. On the other hand, a low success ratio of adversarial attacks in some other applications will not cause severe damage. Moreover, including the misclassification samples in this adaptive design improves the overall detection accuracy of adversarial attacks. The details are explained in the following subsections.

2.3.1. Outputs of Our Proposed Adversarial-Aware Image Recognition System

Our image recognition system has two possible outputs: (1). the image under inspection is authentic, and its *identified* label is provided, or (2). the image under inspection is forged by AML and tagged as *forged*. Therefore, given that there are three possible sets of images in terms of DNN identification (introduced in Section 2.2.1), here are the six possible decision scenarios by our proposed system:

- Decision A ($Dec_a$): An image in SETcrc that is authentic and correctly identified.
- Decision B ($Dec_b$): An image in SETmis that is correctly identified as forged.

---

**Algorithm 2** Adversarial-Aware Deep Learning System.

**[forged, label] = Adv-aware (x)**
**Input:** image x.
**Output:** Whether the image is forged by adversarial attack or clean image; classification label if x is a clean image.

$y \leftarrow DNN(x)$   #*DNN model classification label for the image x*

Top_k $\leftarrow RF(x, k)$   #*The top k group of labels generated by the RF classification model*

**if** $y \in$ Top_k **then**

   forged = false; label = $y$

**else**

   forged = true; label = None

**end if**

**return** [forged, label]

---

- Decision C ($Dec_c$): An image in SETadv that is correctly identified as forged.
- Decision D ($Dec_d$): An image in SETcrc that is misidentified as forged.
- Decision E ($Dec_e$): An image in SETmis that is misidentified as authentic and misclassified.
- Decision F ($Dec_f$): An image in SETadv that is misidentified as authentic.

From a user's perspective, $Dec_a$, $Dec_b$, and $Dec_c$ are all 'good' and rightful decisions; $Dec_d$, $Dec_e$, and $Dec_f$ are wrongful decisions that could cause negative impact/cost to the user.

2.3.2. Adjustable Parameter in Our Proposed System

In our proposed adversarial-aware image recognition system, one critical parameter that can be adjusted/controlled by the end user is the value of *k* in the Top_k classification by the secondary model. It can be used to make a delicate trade-off between increasing the defense accuracy of adversarial attack images and increasing the correct recognition of normal images. The secondary verification ML module determines if an image under inspection belongs to one of the Top_k classes among all possible classification classes. Its classification setting Top_k can be, for example, Top_1, Top_10, Top_20, etc. When *k* increases, the classification decision by the DNN module will have a higher probability of being included in the Top_k classes of the secondary verification system, which will increase the possibility of good $Dec_a$ and increase the possibility of bad Decisions $Dec_e$ and $Dec_f$ as well.

In this paper, we present a solution to the above dilemma by translating and quantifying the problem into optimizing a careful-defined objective cost function. We explain it in detail below.

2.3.3. Using Objective Cost Function to Achieve Optimal Defense

Generally speaking, in most computer vision applications, a successful AML attack will have much more damage to the user than a misclassified event. In most cases, misclassifying an object/content in an image will lead to a clearly identifiable wrongful conclusion, such that the user can easily know that this is a wrong identification. For example, misidentifying a road STOP sign as a red balloon in autonomous vehicle driving will indicate that this is wrong image identification. However, a successful AML attack could make the user misidentify the STOP sign as a SPEED LIMIT sign, which could result in a serious car accident.

For this reason, when we decide how to adjust detection and defense settings for our proposed system, we should not use the classification accuracy, AUC score, or attack success rate directly as the metric. Instead, we define an overall cost objective function, that

**Algorithm 3** Adaptive Design Algorithm.
___
**[k] = Adaptive(DNN classification results, RF classification results)**
**Input:** CIFAR-100(test set), DNN, RF
**Output:** optimal parameter $k$ for the secondary RF model

**for** $k \in \{1, 100\}$ **do**
  Set all the counters $N_a, N_b, \ldots, N_f$ to 0
  **for** image $x \in CIFAR - 100$ **do**
    **if** $x \in SETcrc$ & $DNN(x) \in RF(x, k)$ **then**
      $N_a + +$
    **else**
      $N_d + +$
    **end if**
    **if** $x \in SETmis$ & $DNN(x) \notin RF(x, k)$ **then**
      $N_b + +$
    **else**
      $N_e + +$
    **end if**
    **if** $x \in SETadv$ & $DNN(x) \notin RF(x, k)$ **then**
      $N_c + +$
    **else**
      $N_f + +$
    **end if**
  **end for**
  Objective function $f(k) = (C_d \cdot N_d + C_e \cdot N_e + C_f \cdot N_f - C_a \cdot N_a - C_b \cdot N_b - C_c \cdot N_c)$
**end for**
Among all $f(k), k \in \{1, 100\}$ find the minimum $f(k^*)$
**Return** the optimal index $k^*$
___

is, the weighted summation of all image classification results to find the optimal defense parameters that minimize this objective function.

For the six decision outputs of our proposed system ($Dec_a$ to $Dec_f$), each decision for one image will have its own cost (due to misidentification) or gain (due to correct identification), which can be treated as a positive or a negative costs. Let us define $C_a$, $C_b$, and $C_c$ as the gains for each of those three good decisions $Dec_a$, $Dec_b$ and $Dec_c$; $C_d$, $C_e$, and $C_f$ are the cost values for each of those three wrongful decisions $Dec_d$, $Dec_e$ and $Dec_f$.

The objective cost function $Objf(k)$ for choosing the optimal defense parameter Top_k in the secondary RF classification module is illustrated in Algorithm 3 and shown in Equation 1. We find the optimal value of $k$ by selecting the minimum output ($min_k$) from the equation when changing $k$ from 1 to 100. The parameters $N_a$ to $N_f$ refer to the number of times when decisions $Dec_a$ to $Dec_f$ happen, respectively.

$$Objf(k) = min_k(C_d \cdot N_d + C_e \cdot N_e + C_f \cdot N_f - C_a \cdot N_a - C_b \cdot N_b - C_c \cdot N_c) \quad (1)$$

To calculate $N_a$, $N_b$, ..., and $N_f$, a loop is being conducted over the entire test set of CIFAR-100 dataset. In Algorithm 3, each image x from the dataset is divided into three sets previously from Algorithm 1 (SETcrc, SETmis, and SETadv). Each *if* statement checks whether x image belongs to one of the sets and whether the outcomes of each model prediction (DNN and RF) are matched. For example, if x is a human object and DNN identifies it correctly, and the prediction also exists in the Top_3 RF outcomes, then the decision state is set to $Dec_a$ and $N_a$ counter increments by one.

This optimization is conducted after the training stage when we know the ground truth of all images as shown in Section 3, and can calculate the values of $N_a$ to $N_f$ for each Top_k parameter for all test images. Since the number of possible values of *k* is limited (in our model, it has 100 possible values ranging from 1 to 100), therefore, there is no technical challenge in solving the optimization problem.

2.3.4. Examples of Adjusting Weights on different applications

In this section, we use several application scenarios to show why they need different cost weights in our adaptive design and the above optimization Equation 1. In different image classification applications, users can define the concrete values for the other cost factors according to their expert opinion and application scenarios. Four applications are introduced in the following, and Table 4 in the following section presents the outcomes of this adaptive method.

- **Autonomous driving**: we can define $C_a$ = 0.3, $C_b$ = 0.1, and $C_c$ = 0.5. The value of $C_c$ is higher than $C_a$ because, in autonomous driving, it is more important for us to detect an adversarial attack than correctly identify a normal roadside sign image. Similarly, we can define $C_d$ = 0.1, $C_e$ = 0.3, and $C_f$ = 0.8. We define $C_f$ as having a significantly higher value than others because $Dec_f$ means autonomous driving is compromised under a deliberate adversarial attack. For example, we could treat a STOP sign image as a right-turn-only sign, which could result in serious accident consequences. The value of $C_e$ is higher than $C_b$ in detecting misclassified images by the model due to the risk value we assume.
- **Healthcare**: although deep learning based healthcare systems could achieve high accuracy in disease diagnosis, few such systems have been deployed in highly automated disease screening settings due to a lack of trust. Therefore, the human-based double-check process is usually used, and hence, the deep learning healthcare system can be tolerated in the security. An example values of the weights are $C_a$ = 0.7, $C_b$ = 0.4, $C_c$ = 0.1, $C_d$ = 0.4, $C_e$ = 0.1, and $C_f$ 0.3. The $C_a$ is the highest cost weight because the physician will most likely discover failure in other decisions during manual double-checking.
- **Face recognition in checking work attendance**: misrecognition or adversarial impact is low because the potential of utilizing these challenges by the employees is rare. Therefore, we can give the positive gains higher values with $C_a$ = 0.7, $C_b$ = 0.4, and $C_c$ = 0.2. In contrast, we can value the negative decisions as $C_d$ = 0.4, $C_e$ = 0.2, and $C_f$ = 0.2.
- **Detecting inappropriate digital content**: mispredicting nudity images to protect children is another example where the costs of AML attack are medium, not as risky as in autonomous driving, nor as tolerable as in face recognition. Hence, we can choose $C_a$ = 0.7, $C_b$ = 0.1, $C_c$ = 0.2, $C_d$ = 0.3, $C_e$ = 0.1, and $C_f$ = 0.1.

2.3.5. The Cost of Misclassified Clean Images

As far as of today, there are no image classification models that can provide a 100 percent accurate result. Table 1 shows the accuracy rates of various DNN models without any attacks. ResNet-34 achieves an accuracy rate of 77.47 percent, while VGG16 has a lower accuracy rate of 72.25 percent. On the other hand, DenseNet boasts a higher accuracy rate of 78.69 percent. The percentage of misclassified images is enormous. Therefore, the business models of AI applications should consider these failure cases to assess their risks in case of using any DNN model with a high percentage of misclassification. On the other hand, our approach can detect a significant fraction of these detection failures and categorize them as forged by adversarial attacks.

As described in the previous section, $Dec_b$ can identify the misclassification of tested samples and be counted as positive to DNN model accuracy. On the other hand, $Dec_e$, where our approach wrongly identifies it as forged, is counted as negative to the overall accuracy. Application designers can define the costs of these decisions, balancing security and safety with passing tolerance using Algorithm 3. The accuracy of the overall system can be significantly affected, as demonstrated in Table 4 in Section 3.

2.3.6. Evaluation Metric

The evaluation technique for our proposed method is similar to previous works of detection methods in [9–11]. We use the Area under the ROC Curve (AUC) score in our assessments between clean $Dec_a$ and adversarial images $Dec_c$, as it will be addressed in Section 3 at Table 3. Accuracy (acc.) is another metric used to evaluate our proposed model based on image classification application parameters that will be introduced in Section 2.3 and Section 3.

## 3. Results

In this section, we showcase the evaluation and outcomes of our study. First, settings for the experiments and the environment utilized are explained. Then, the configurations for the adversarial attacks we deploy to target the various deep learning models are outlined. Lastly, we present and compare the main results according to each proposed approach in sections 2.2 and 2.3.

*3.1. Experimental Setup*

To evaluate the robustness and effectiveness of the proposed scheme, we run our training, evaluation, and attacks using NVIDIA GeForce RTX 3090 GPU. We use Sklearn [36] open-source Python library for the classical ML Random Forest model. On the other hand, we use PyTorch-lightning [37] for DNN models. Finally, we use Torchattacks [38] to run the adversarial attacks.

*3.2. Adversarial Attack Configuration*

The attacker knows that the targeted image classification system uses ResNet-34 for training the image classification model. He/she also knows the data being used for that training which is the CIFAR-100 training set. The attacker will use a test set of the same dataset and state-of-art adversarial attack algorithms: FGSM [12], Deepfool [13], CW [14], and PGD [15]. The parameters of each type of AML are listed in Table 2 and defined in the next.

**Table 2.** Experiment settings.

| Targeted Model | Dataset | Adversarial Attacks | Parameters | Attack Success Ratio (%) |
|---|---|---|---|---|
| ResNet-34 | CIFAR-100 | FGSM | $\epsilon$ = 0.007 | 65.75 |
| | | Deepfool | s = 50, overshoot = 0.02 | 99.92 |
| | | CW | c = 1.0, $\kappa$ = 0, s = 50, lr = 0.01 | 98.64 |
| | | PGD | $\epsilon$ = 0.03, $\alpha$ = 0.004, s = 40 | 98.83 |

In the FGSM trial, we set the parameter $\epsilon$ at 0.007, which is a hyperparameter determining the size of the perturbations introduced to the input data. The value of $\epsilon$ is a trade-off between the adversarial attack strength and the perturbation perceptibility. Raising this value could increase the exploit success rate; however, it might show an apparent noise on the targeted image that could be revealed to human perception. We select the default value as 0.007 because the added perturbations are not easily perceived by human eyes. The FGSM attack success accuracy based on the selected $\epsilon$ on the CIFAR-100 test set is 65.75%.

To execute the Deepfool attack, we limit the attack iterations to 50 steps before stopping. During each iteration, the attack calculates the direction of the closest decision boundary to the original input data point in order to determine the minimum perturbation required to deceive the targeted DNN model. The overshoot parameter is set to 0.02, which multiplies the computed perturbation vector and adds it to the input image. With these settings, the attack success accuracy reaches 99.92%.

To ensure a successful attack by CW method, we utilize the C&W attack parameters listed in Table 2: $c = 1$, $\kappa = 0$, steps $s = 50$, and lr = 0.01. The '$c$' hyperparameter determines the magnitude of the perturbation, while the margin parameter $\kappa$ determines the confidence gap between the predicted and target classes. The steps $s$ parameter represents the number of iterations required for the attack to succeed or end. Lastly, the learning rate 'lr' controls the optimization iteration steps. With these adjustments, we achieve an attack success rate of 98.64%.

The PGD attack was adjusted with the following parameters: $\epsilon = 0.03$, alpha $\alpha = 0.004$, and steps = 40. $\epsilon$ and steps were explained in previous attacks, while alpha functions similar to the learning rate, determining the size of each optimization step. This attack has a success rate of 98.83%.

*3.3. Main Results*

Tabel 3 summarizes the AUC scores of four adversarial attack detectors with our proposed method from Section 2.2 using features from all the DNN penultimate layers. For comparison, we compare our proposed method with four other popular adversarial detection methods, DkNN [8], LID [9], Mahalanibis [10], and NNIF [11].

Overall, our proposed Top_1 threshold surpasses other methods in most attacks, as indicated in bold, while it founds that the LID method was the least effective in detecting attacks. The best FGSM attack detection goes to our proposed Top_22 method. Additionally, the NNIF model is the second-best detector approach to resist all attacks. The AUC scores at FGSM show a roughly 10 percent gap between the detectors. In contrast, in Deepfool, the gap was much more pronounced, with LID scoring 52.25 and our proposed Top_1 scoring 97.57.

The AUC score comparisons for different adversarial detector models on various attacks are shown in Figure 3. The x-axis represents the four adversarial attacks, while the y-axis describes the AUC score ranging from zero to 100. Each color on the graph represents one defense method as represented in the top right legend, namely DkNN, LID, Mahalanibis, NNIF, proposed prpo[T1], and proposed prpo[T22] with colors grey, navy, light green, light pink, light blue, and light brown, respectively. In the FGSM attack, DkNN and proposed (prpo[T22]) were the most effective defense mechanisms, while the others showed slight differences with a score of around 80. The Deepfool bars showed significant improvement in detection methods, but some methods had noticeable weaknesses. For the CW attack detection, DkNN, NNIF, and proposed (T1) performed well, while the others scored an average of 70. Finally, the PGD attack proved to be powerful against DkNN, LID, and Mahalanibis with a semi-match AUC score of 72, while the remaining methods saw significant improvement, with proposed prpo[T1] scoring the highest with a score of 96.

**Table 3.** AUC Score of Adversarial Detection Methods.

| Detectors | AUC Score | | | |
|---|---|---|---|---|
| | FGSM | Deepfool | CW | PGD |
| DkNN [8] | 93.65 | 76.71 | 93.77 | 73.78 |
| LID [9] | 80.68 | 52.25 | 67.84 | 72.25 |
| Mahalanibis [10] | 83.90 | 62.05 | 71.60 | 72.46 |
| NNIF [11] | 87.23 | 84.20 | 94.58 | 83.09 |
| Top_1 | 86.62 | **97.57** | **98.21** | **96.49** |
| Top_22 | **94.17** | 74.17 | 83.50 | 86.04 |

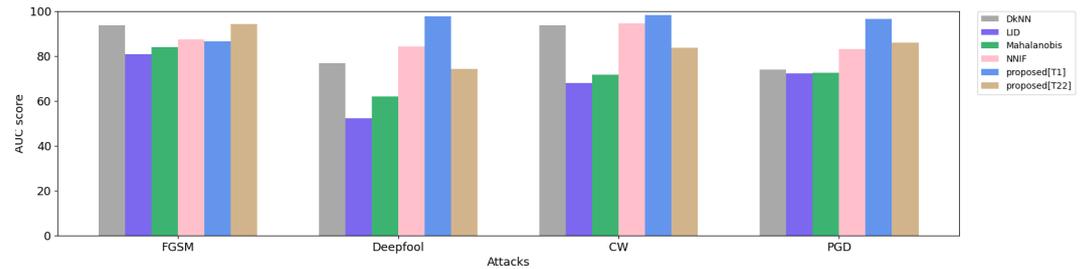

**Figure 3.** AUC score comparison for adversarial attacks detectors. The x-axis represents the detector methods. The y-axis represents the AUC score of adversarial detectors. Each color demonstrates one of the detectors as it is listed in the right top legend.

For our proposed system, there is an inherent trade-off between higher accuracy in detecting adversarial attacks and higher classification accuracy for clean data inputs, as illustrated in Table 4. We assign varying weights to each application depending on the potential risks we might face in the event of overlooking a successful attack and also depending on how accurate we prefer in classifying normal clean inputs. Our adaptive optimization algorithm (Equation 1) has determined that the optimal settings for RF Top_$x$ probability index should be as follows. In all types of attacks, autonomous driving takes the Top_1 due to the potential for severe accidents if adversarial or misclassified samples are not detected. In healthcare, FGSM takes Top_5, and the remaining attacks all take Top_1. Face Recognition application selects Top_3 for FGSM and Top_1 for the rest. Finally, detecting inappropriate content on a system selects Top_14 for the FGSM attack and Top_1 for the other attacks.

In Section 2.3.5, we discussed how misclassification samples could improve the accuracy of the ResNet-34 model in detecting adversarial attacks. To demonstrate this, we conduct an FGSM attack experiment using the same applications and weights as in Table 4. We present the results in Table 5. First, we calculate the accuracy without the misclassification samples, using Equation 1. Then, we calculate the accuracy again after including misclassification samples ($C_b \cdot N_b$ and $C_e \cdot N_e$), as displayed in Table 5. Our approach effectively enhances the detecting AML accuracy on ResNet-34 model, which was initially predicted with 74.47 percent accuracy.

**Table 4.** AUC score comparison based on different applications preferences.

| Applications | Wights | Accuracy based on best Top_n selection from formula (1) | | | | | | | |
|---|---|---|---|---|---|---|---|---|---|
| | | FGSM | acc. | DeepFool | acc. | CW | acc. | PGD | acc. |
| Autonomous driving | $C_a = 0.3$<br>$C_b = 0.1$<br>$C_c = 0.5$<br>$C_d = 0.1$<br>$C_e = 0.3$<br>$C_f = 0.8$ | Top_1 | 81.14% | Top_1 | 89.84% | Top_1 | 89.68% | Top_1 | 90.04% |
| Healthcare | $C_a = 0.7$<br>$C_b = 0.4$<br>$C_c = 0.1$<br>$C_d = 0.4$<br>$C_e = 0.1$<br>$C_f = 0.3$ | Top_5 | 77.66% | Top_1 | 89.84% | Top_1 | 89.68% | Top_1 | 90.04% |
| Face Recognition | $C_a = 0.7$<br>$C_b = 0.4$<br>$C_c = 0.2$<br>$C_d = 0.4$<br>$C_e = 0.2$<br>$C_f = 0.2$ | Top_3 | 79.06% | Top_1 | 89.84% | Top_1 | 89.68% | Top_1 | 90.04% |
| Inappropriate content | $C_a = 0.7$<br>$C_b = 0.1$<br>$C_c = 0.2$<br>$C_d = 0.3$<br>$C_e = 0.1$<br>$C_f = 0.1$ | Top_14 | 70.14% | Top_1 | 89.84% | Top_1 | 89.68% | Top_1 | 90.04% |

**Table 5.** AML Detection Accuracy comparison before and after including misclassification samples

| Applications | w/o misclassification (%) | with misclassification (%) |
|---|---|---|
| Autonomous driving | 62.81 | 81.14 |
| Healthcare | 63.20 | 77.66 |
| Face Recognition | 61.66 | 79.60 |
| Detecting inappropriate content | 60.08 | 70.14 |

## 4. Discussion

This section links our proposed ideas with the results and provides a more insightful summary and discussion. First, we compare our model with others from a similar family (external detectors). After that, we discuss the results from Tables 1, 2, 3, 4, and 5. We then express our challenges and limitations.

Even though the RF is a sufficient model in regression [30] and multi-classification applications [39], it is not commonly used for image classifications because images have a large number of pixels, resulting in high-dimensional feature spaces. In addition, image processing is computationally expensive and time-consuming during training. However, we have decided to use RF as a secondary model for two reasons. Firstly, other models like Support Vector Machine (SVM) [40] are not efficient in multi-classifications and are computationally expensive. Secondly, we want to showcase the usefulness of having two different architectural models to overcome adversarial attacks. Studies such as [8,11] have used traditional ML algorithms to create AML detectors. They adapted *k*-NN in generating their adversarial detectors by adding *k*-NN model between DNN layers during training to extract new features that can be analyzed and used to recognize clean and noisy samples versus adversarial ones. However, besides *k*-NN's extreme complexity and high computational performance, these studies have found that different types of attacks have varying resistances, depending on the effectiveness of the attack in generating perturbations to fool the model.

For example, Table 3 shows fluctuations in AUC scores in resisting each attack by every detection algorithm, such as NNIF for FGSM, Deepfool, CW, and PGD, with scores of 87.23, 84.20, 94.58, and 83.09, respectively. In contrast, our proposed system with Top_1 setting is consistently effective regardless of FGSM outcomes, as it has a large number of clear samples that are not affected by the attack at $α$ 0.007; further clarification will be provided lately. Therefore, a new technique of attack that relies on backpropagation could harden the defense algorithms as illustrated in [11] when trained detector on FGSM attack only tested on unseen attacks such as Deepfool. Their findings indicated a decline in performance when testing for unseen attacks compared to seen attacks. Our proposed system, however, is tested on all adversarial attacks without attack patterns evaluation nor DNN models changing and presents a generalization across different attacks.

The results of the FGSM attack in Table 4 show reasonable changes at selecting Top_$k$ based on an application's weight parameters. Unlike the other attacks, our defense system remains fixed with Top_1. This change from Top_1 in autonomous driving to Top_14 in detecting inappropriate content is normal when we increase the cost of $C_f$. In this situation, Equation 1 significantly increases the security sensor to minimize the success rate of adversarial attacks. In contrast, the equation reduces the model sensitivity when preventing inappropriate content because the risk of successful attacks is not so serious. This equation provides freedom to the application developer to choose the best and most optimal defense setup.

Our approach of using Top_$k$ has bridged the gap in efficiency between DNN, which is a model for image classification, and the RF model, which is suitable for structured and tabular data. It also allows for a choice between security and identification, as explained in Section 2.3. However, Table 4 shows that we consistently use Top_1 for every test of Deepfool, CW, and PGD attacks because the success rate of these attacks, as shown in Table 2, is unrealistic. In real-world attacks, attackers do not have access to information about the ML models or the data used for training, the level of security integrated, etc. The regular successful rate should be much less than what we deal with in the presented examples in the table. FGSM attack shows an example of the successful usage of our proposed adaptive design theory; otherwise, the system's adaptivity is useless if applied systems are extremely exposed to adversarial attacks.

## 5. Conclusions

Our paper presents a straightforward yet impactful approach to identifying adversarial attacks. It involves utilizing a secondary classical machine learning model in conjunction with the primary DNN model for image classification. The secondary model's architecture completely differs from the primary model to thwart adversarial attacks that rely on the backpropagation technique that generates an effective perturbation targeting the primary deep learning model. Our proposed detector outperforms state-of-the-art models that rely on analyzing adversarial sample behavior and patterns during DNN training. Additionally, our model requires no modifications to DNN model or learning attack types. We use CIFAR-100 dataset for this study as it contains reasonable number of classes to fulfill this task considering all the challenges with RF model in image classifications tasks.

Challenges are found throughout this study. Foremost, the RF model is designed professionally for structural and tabular applications such as stock market price predictions that use a specific number of vectors, simple image classification, or recognition tasks like satellite imagery object detection. However, the RF model capability can be limited when faced with large-scale training involving a significant number of classes, like when using the ImageNet dataset [41] with its 1000 classes and 1.2 million images. To tackle these challenges, a DNN model was developed. It excels in extracting features from high-dimensional vector spaces and large datasets while requiring less time for training. We use this RF model as a prototype for our analysis, and we highlight that a potential robust defense mechanism exists if we can adopt a different architecture.

In our future work, we aim to improve our detector by dealing with the scalability issue in classical ML models. To achieve this, we plan to use feature vectors from DNN fully connected layers and input the outputs to RF. This strategy will create a more effective detector using RF as a classifier and DNN as a feature extractor. By combining the strengths of both models, we can benefit from DNN's superior feature extraction and RF's outstanding ability to mitigate adversarial attacks.

**Author Contributions:** Conceptualization, M.A. and C.Z.; methodology, M.A.; software, M.A.; validation, M.A, and C.Z.; formal analysis, M.A. and C.Z.; investigation, M.A.; resources, M.A.; data curation, M.A.; writing—original draft preparation, M.A.; writing—review and editing, H.K., M.A., A.A., and C.Z.; visualization, M.A.; supervision, C.Z.; project administration, M.A.; funding acquisition, N.A. All authors have read and agreed to the published version of the manuscript.

**Conflicts of Interest:** The authors declare no conflict of interest.

## Abbreviations

The following abbreviations are used in this manuscript:

| | |
|---|---|
| DNN | Deep Neural Network |
| ML | Machine Learning |
| Top_k | The index of the top k classes of a model prediction |
| AML | Adversarial Machine Learning |
| AI | Artificial Intelligence |
| FGSM | Fast Gradient Sign Method |
| CW | The Carlini and Wagner attack |
| PGD | Projected Gradient Descent attack |
| ComCNN | Compression Convolutional Neural Network |
| k-NN | K Nearest Neighbours |
| LID | Local Intrinsic Dimensionality |
| MART | Misclassification Aware adveRsarial Trainin |
| AUC | Area under the ROC Curve |
| SVM | Support Vector Machine |